\documentclass[12pt]{article}
\usepackage{graphicx}
\usepackage{amsmath}
\usepackage{amsbsy}
\usepackage{cite}
\usepackage{slashed}
\usepackage{amsfonts}
\usepackage{amssymb}

\newcommand\pubnumber{LPSC14223}
\newcommand\pubdate{\today}

\def\Title#1{\begin{center} {\Large #1 } \end{center}}
\def\Author#1{\begin{center}{ \sc #1} \end{center}}
\def\Address#1{\begin{center}{ \it #1} \end{center}}

\newcommand\pubblock{\rightline{\begin{tabular}{r} \pubnumber\\
         \pubdate  \end{tabular}}}
\newenvironment{Abstract}{\begin{quotation}  }{\end{quotation}}
\newenvironment{Presented}{\begin{quotation} \begin{center} 
             PRESENTED AT\end{center}\bigskip 
      \begin{center}\begin{large}}{\end{large}\end{center} \end{quotation}}

\def\beq{\begin{equation}}
\def\eeq#1{\label{#1}\end{equation}}
\def\eeqn{\end{equation}}

\def\beqa{\begin{eqnarray}}
\def\eeqa#1{\label{#1}\end{eqnarray}}
\def\eeqan{\end{eqnarray}}

\let\bar=\overbar

\def\Dslash{\not{\hbox{\kern-4pt $D$}}}
\def\dslash{\not{\hbox{\kern-2pt $\del$}}}

\def\msb{{\bar{\ssstyle M \kern -1pt S}}}

\begin{document}
\begin{titlepage}
\pubblock

\vfill
\Title{Rare K decays: Challenges and Perspectives}
\vfill
\Author{Christopher Smith}
\Address{Laboratoire de Physique Subatomique et de Cosmologie, Universit\'{e} Grenoble-Alpes, CNRS/IN2P3, 53 avenue des Martyrs, 38026 Grenoble Cedex, France}
\vfill
\begin{Abstract}
At this stage of the LHC program, the prospect for a new physics signal in the very rare $K\rightarrow\pi\nu\bar{\nu}$ decays may be dented, but remains well alive thanks to their intrinsic qualities. First, these decays are among the cleanest observables in the quark flavor sector. When combined with their terrible suppression in the SM, they thus offer uniquely sensitive probes. Second, the LHC capabilities are not ideal for all kinds of new physics, even below the TeV scale. For example, rather elusive scenarios like natural-SUSY-like hierarchical spectrum, baryon number violation, or new very light but very weakly interacting particles may well induce deviations in rare $K$ decays. Even though experimentalists should brace themselves for tiny deviations, these modes thus have a clear role to play in the LHC era.
\end{Abstract}
\vfill
\begin{Presented}
Flavor Physics and CP Violation (FPCP-2014)\\Marseille, France, May 26--30, 2014
\end{Presented}
\vfill
\end{titlepage}

\section{Why studying rare $K$ decays in the LHC era?}

Most models of new physics (NP) introduce either new flavored particles, or new flavor-breaking interactions between quarks and leptons. Even if these new particles are relatively heavy, their presence can alter the delicate balance at play in the Flavor Changing Neutral Currents (FCNC). Indeed, those processes would actually vanish at all orders in the Standard Model (SM) if quark masses were equal because of the unitarity of the CKM matrix, the so-called GIM mechanism. So, FCNC arise in the SM from a loop-level interplay between the non-degeneracy of quark masses and the non-diagonal nature of the CKM matrix~\cite{BuchallaBL96}. On the other hand, the NP dynamics may not be so contrived, and could in principle directly lead to FCNC.

Let us be a bit more precise. Concentrating on $K$ and $B$ semileptonic decays with neutrino pairs in the final states, both the SM and the NP contributions can be embodied into dimension-six effective operators%
\begin{equation}
\mathcal{H}_{eff}=\frac{c^{bs}}{\Lambda^{2}}\,\bar{b}\Gamma s\otimes\bar{\nu}\Gamma\nu
                 +\frac{c^{bd}}{\Lambda^{2}}\,\bar{b}\Gamma d\otimes\bar{\nu}\Gamma\nu
                 +\frac{c^{sd}}{\Lambda^{2}}\,\bar{s}\Gamma d\otimes\bar{\nu}\Gamma\nu+...
\end{equation}
where $\Gamma$ represents some Dirac structures, $\Lambda$ is the typical scale of the dynamics, and $c^{ij}$ are the Wilson coefficients. With this language, the absence of NP signals in flavor experiments would mean that $c_{\mathrm{NP}}^{ij}/\Lambda_{\mathrm{NP}}^{2}\lesssim c_{\mathrm{SM}}^{ij}/\Lambda_{\mathrm{SM}}^{2}$, i.e., that the NP contributions are smaller than those of the SM, for which $\Lambda_{\mathrm{SM}}\sim v_{EW}\approx 246$~GeV and $c_{\mathrm{SM}}^{ij}\sim(g^{2}/4\pi)^{2}V_{ti}V_{tj}^{\dagger}$ with $g$ the electroweak coupling and $V_{ij}$ the CKM matrix elements. These combinations of CKM matrix elements scale as%
\begin{equation}
c_{\mathrm{SM}}^{bs}\sim V_{tb}V_{ts}^{\dagger}\sim\lambda^{2}\;,\;\;
c_{\mathrm{SM}}^{bd}\sim V_{tb}V_{td}^{\dagger}\sim\lambda^{3}\;,\;\;
c_{\mathrm{SM}}^{sd}\sim V_{ts}V_{td}^{\dagger}\sim\lambda^{5}\;,
\end{equation}
where $\lambda\sim0.2$ is the sine of the Cabibbo angle. There are several ways to interpret the consequences of these scalings on the search for NP in $K$ decays:

\begin{enumerate}
\item If the NP coefficients $c_{\mathrm{NP}}^{ij}$ do not follow any particular scaling but are all of $\mathcal{O}(1)$, the constraints from the kaon sectors are clearly the tightest. The absence of a NP signal there forces its contribution not to exceed a very small SM piece. Typically, a measurement of $K\rightarrow\pi\nu\bar{\nu}$ compatible with the SM pushes $\Lambda_{\mathrm{NP}}\gtrsim100$ TeV~\cite{BurasZp}. For such a scale, the corresponding impact in $B$ transitions is completely negligible, while the LHC is not energetic enough for a direct discovery. Kaon physics then represent our only window. At the same time, such a scenario is not very appealing theoretically. When $\Lambda_{\mathrm{NP}}\gtrsim10^{4}\times\Lambda_{\mathrm{SM}}$, the stability of the electroweak scale necessitates systematic and extreme fine-tuning of the Lagrangian parameters. This is called the hierarchy puzzle.

\item If $\Lambda_{\mathrm{NP}}$ is lower, as suggested to alleviate the hierarchy puzzle, then another puzzle arises: the $c_{\mathrm{NP}}^{ij}$ have to be suppressed. In that respect, kaon physics is certainly the most puzzling, since $c_{\mathrm{NP}}^{sd}$ should be particularly small. Phenomenologically, a simple way to put back $K$ and $B$ physics on a similar footing is to enforce Minimal Flavor Violation (MFV), i.e., to ensure that $c_{\mathrm{NP}}^{ij}\sim V_{ti}V_{tj}^{\dagger}$~\cite{MFV}. With such a suppression, the NP contributions would be similar to those of the SM if it arises at the loop level and $\Lambda_{\mathrm{NP}}\approx v_{EW}$. So, even a moderately-higher scale would not induce large deviations in flavor observables. Indeed, the NP decoupling is rather fast since the effective operators have dimension six.

\item Direct searches for new particles at the LHC have pushed the typical mass scale for many scenarios close to or even above 1 TeV. But, if $\Lambda_{\mathrm{NP}}\approx5\times v_{EW}$, a loop-level NP contributions is about 25 times smaller than the SM contributions when MFV is active. While this is certainly compatible with the current absence of deviations with respect to the SM in $B$ physics and $K-\bar{K}$ mixing, it raises the issue of the very observability of NP in low-energy experiments~\cite{BurasG14}. A significant signal would require either significant departures from MFV in the $K$ sector, or a much lighter relevant dynamics which would have somehow escaped detection at the LHC, or very specific observables for which the theoretical control on the hadronic uncertainties can be achieved at the few percent level. This is where the very clean rare $K$ decays could play a crucial role.
\end{enumerate}

With this picture in mind, we will concentrate in the following on three different topics. First, in Section 2, we will discuss the hadronic uncertainties playing a role for the rare $K$ decay rate predictions. Since the deviations from the SM predictions are expected to be small for a large class of NP scenarios, particular emphasis will be laid on the methods used to control them, and the prospect for further improvements. In Section 3, we will illustrate the tension between the LHC bounds on NP particle masses and the observability of a deviation in the rare $K$ decay rates in the context of supersymmetric models. Besides the intrinsic appeal of this class of scenarios, this choice is particularly convenient for our purpose since it offers a fully dynamical setting in which the interplay between low- and high-energy observables can be consistently analyzed. We will show that two specific scenarios, natural SUSY\ and R-parity violating SUSY could still manage to lead to significant deviations. Finally, in Section 4, the sensitivity of rare $K$ decays to the presence of new light, neutral, and very weakly interacting particles will be briefly explored.

\section{Controlling hadronic effects\label{HadEff}}

One of the main virtues of the rare $K$ decays is the excellent control achieved over the hadronic effects on their rates, even though the typical scale of kaon physics is deep in the QCD non-perturbative regime. This could be accomplished thanks to the powerful chiral symmetry, which permits to recover some form of perturbative treatment for the strong interaction at low energy. Before illustrating this, let us first identify precisely the quantities affected by hadronic uncertainties.

The dominant contribution to $K\rightarrow\pi\nu\nu$ comes from the $Z$ penguin~\cite{InamiLim}, where the three up-type quarks circulate. It is CP-conserving for $K_{1}\rightarrow\pi^{0}\nu\bar{\nu}$ and CP-violating for $K_{2}\rightarrow\pi^{0}\nu\bar{\nu}$, where $\sqrt{2}K_{1,2}=K^{0}\mp\bar{K}^{0}$ are the $0^{++}$ and $0^{-+}$ neutral kaon CP-eigenstates, approximately equal to the mass eigenstates $K_{S}$ and $K_{L}$, respectively. The loop function induces a quadratic breaking of the GIM mechanism~\cite{GIM}, i.e., it is proportional to $m_{q}^{2}/M_{W}^{2}$ in the $m_{q}\rightarrow\infty$ and $m_{q}\rightarrow0$ limit, where $m_{q}$ the mass of the quark circulating in the loop. Combined with the CKM scaling for the CP-conserving and CP-violating transitions, we get:%
\[%
\begin{tabular}[c]{ccc}
& \multicolumn{2}{c}{$K^{+}\rightarrow\pi^{+}\nu\bar{\nu}$}\\
& $K_{1}\rightarrow\pi^{0}\nu\bar{\nu}\;\;$ & $K_{2}\rightarrow\pi^{0}\nu\bar{\nu}$\\
\raisebox{-0.27in}{\includegraphics[height=0.65in]{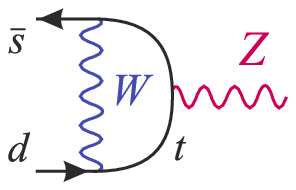}}$\;\;\;\;\;$ & 
$\dfrac{m_{t}^{2}}{M_{W}^{2}}(\operatorname{Re}V_{ts}^{\dagger}V_{td}\sim\lambda^{5})\;\;\;\;\;$ & $\dfrac{m_{t}^{2}}{M_{W}^{2}}(\operatorname{Im}V_{ts}^{\dagger}V_{td}\sim\lambda^{5})$\\
\raisebox{-0.27in}{\includegraphics[height=0.65in]{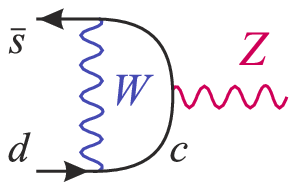}} $\;\;\;\;\;$ & 
$\dfrac{m_{c}^{2}}{M_{W}^{2}}(\operatorname{Re}V_{cs}^{\dagger}V_{cd}\sim\lambda)\;\;\;\;\;$ & 
$\dfrac{m_{c}^{2}}{M_{W}^{2}}(\operatorname{Im}V_{cs}^{\dagger}V_{cd}\sim\lambda^{5})$\\%
\raisebox{-0.27in}{\includegraphics[height=0.65in]{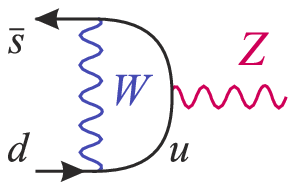}}$\;\;\;\;\;$ & 
$\dfrac{m_{u}^{2}}{M_{W}^{2}}(\operatorname{Re}V_{us}^{\dagger}V_{ud}\sim\lambda)\;\;\;\;\;$ & 
$\dfrac{m_{u}^{2}}{M_{W}^{2}}(\operatorname{Im}V_{us}^{\dagger}V_{ud}=0)$%
\end{tabular}
\ \ \ \ \ 
\]
The $K^{+}\rightarrow\pi^{+}\nu\bar{\nu}$ decay mode receives both a CP-conserving and a CP-violating contribution since $K^{+}$ is not a CP eigenstate.

These scalings explain why the top quark contribution is so large both for the CP-violating and CP-conserving transitions. The purely long-distance up quark contribution~\cite{kpnnLD,IsidoriMS05} is suppressed by the light quark mass, and is necessarily CP-conserving. The charm quark contribution ends up as large as the top one for the CP-conserving transition, because the small mass ratio $m_{c}^{2}/m_{t}^{2}$ is compensated by the large CKM ratio $\operatorname{Re}V_{cs}^{\dagger}V_{cd}/\operatorname{Re}V_{ts}^{\dagger}V_{td}\sim\lambda^{-4}$, but stays subleading for the CP-violating transition~\cite{kpnnSD,BGS11}. Note that indirectly, this large CP-conserving charm quark contribution contributes to the $K_{L}\rightarrow\pi^{0}\nu\bar{\nu}$ decay. Indeed, the kaon mass eigenstates are $K_{S}\sim K_{1}+\varepsilon K_{2}$ and $K_{L}\sim K_{2}+\varepsilon K_{1}$. Thankfully, $\varepsilon\sim10^{-3}$, so this so-called indirect CP-violating piece enters only at the percent level in the $K_{L}\rightarrow\pi^{0}\nu\bar{\nu}$ rate~\cite{BB96}. Finally, it must be mentioned that if a different CKM phase convention was chosen, for example one in which $\operatorname{Im}V_{us}^{\dagger}V_{ud}\neq0$, then it is only through the interference of the direct and indirect CP-violating amplitudes that these scalings between the three up-type quark contributions would be recovered~\cite{GrossmanN97}.

To really appreciate how peculiar is the $Z$ penguin, it is instructive to compare with the photon penguin, for which the GIM breaking is logarithmic:%
\[%
\begin{tabular}[c]{ccc}
& \multicolumn{2}{c}{$K^{+}\rightarrow\pi^{+}\gamma^{\ast}$}\\
& $K_{1}\rightarrow\pi^{0}\gamma^{\ast}\;\;\;$ & $K_{2}\rightarrow\pi^{0}\gamma^{\ast}$\\%
\raisebox{-0.27in}{\includegraphics[height=0.65in]{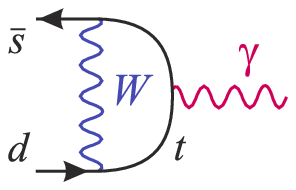}}$\;\;\;\;\;$ & 
$\log\dfrac{m_{t}}{M_{W}}(\operatorname{Re}V_{ts}^{\dagger}V_{td}\sim\lambda^{5})\;\;\;\;\;$ & 
$\log\dfrac{m_{t}}{M_{W}}(\operatorname{Im}V_{ts}^{\dagger}V_{td}\sim\lambda^{5})$\\%
\raisebox{-0.27in}{\includegraphics[height=0.65in]{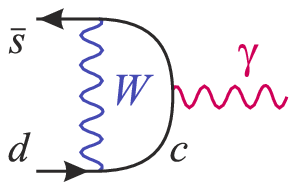}}$\;\;\;\;\;$ & 
$\log\dfrac{m_{c}}{M_{W}}(\operatorname{Re}V_{cs}^{\dagger}V_{cd}\sim\lambda)\;\;\;\;\;$ & 
$\log\dfrac{m_{c}}{M_{W}}(\operatorname{Im}V_{cs}^{\dagger}V_{cd}\sim\lambda^{5})$\\%
\raisebox{-0.27in}{\includegraphics[height=0.65in]{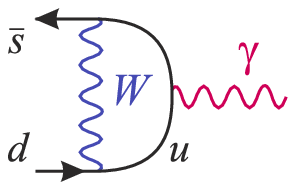}}$\;\;\;\;\;$ & 
$\log\dfrac{m_{u}}{M_{W}}(\operatorname{Re}V_{us}^{\dagger}V_{ud}\sim\lambda)\;\;\;\;\;$ & 
$\log\dfrac{m_{u}}{M_{W}}(\operatorname{Im}V_{us}^{\dagger}V_{ud}=0)$%
\end{tabular}
\ \ \ \ \
\]
This time, the CP-conserving up-quark contribution is not suppressed at all. Actually, it dominates the CP-conserving process, relegating the top contribution to a negligible correction~\cite{DEIP98}. As a result, neither $K_{S}\rightarrow\pi^{0}\ell^{+}\ell^{-}$ nor $K^{+}\rightarrow\pi^{+}\ell^{+}\ell^{-}$ qualify as good probes for the short-distance dynamics. On the other hand, $K_{L}\rightarrow\pi^{0}\ell^{+}\ell^{-}$ still shares many of the good features of $K\rightarrow\pi\nu\bar{\nu}$. One must be careful though to treat the indirect CP-violating contribution~\cite{BDI03}, which scales as $\varepsilon\times\log m_{u}/M_{W}\times\operatorname{Re}V_{us}^{\dagger}V_{ud}$. The long distance enhancement is so strong that it completely compensate for the small $\varepsilon$, and this contribution ends up of the same order as the CP-violating top and charm quark contributions (in other words, the so-called ``indirect'' and ``direct'' CP violating contributions are similar). Finally, the two-photon penguin contribution $K_{2}\rightarrow\pi^{0}\gamma^{\ast}\gamma^{\ast}\rightarrow\pi^{0}\ell^{+}\ell^{-}$ should also be included. Being CP-conserving, $K_{2}\rightarrow\pi^{0}\gamma^{\ast}\gamma^{\ast}$ is also entirely dominated by the up-quark contribution and enhanced by the long-distance meson dynamics. The situation is different when coupled to muons or electrons. The two photons can have $J^{CP}=0^{++},2^{++}$,... , but mostly the scalar state is produced. As a result, the lepton amplitude is helicity-suppressed (proportional to the lepton mass). The two-photon contribution is then comparable to the direct and indirect CP-violating contribution for the muon mode~\cite{ISU04}, but remains negligible for the electron one~\cite{BDI03}.

The electroweak anatomy of the various decay processes permits first to identify the most promising decays, i.e., those most sensitive to the interesting short-distance dynamics: $K_{L}\rightarrow\pi^{0}\nu\bar{\nu}$, $K^{+}\rightarrow\pi^{+}\nu\bar{\nu}$, and to a lesser extent, $K_{L}\rightarrow\pi^{0}e^{+}e^{-}$ and $K_{L}\rightarrow\pi^{0}\mu^{+}\mu^{-}$. Then, it also permits to identify two sources of hadronic effects. In both cases, the strategy to control these effects is to use the chiral symmetry to relate them to well-measured quantities. Specifically:

\begin{figure}[t]
\centering                
\includegraphics[width=10.4cm]{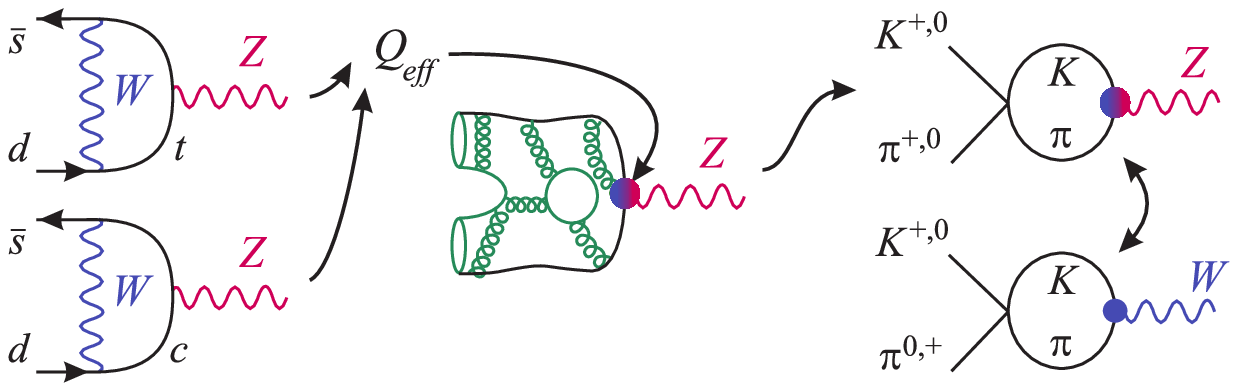}
\includegraphics[width=10.4cm]{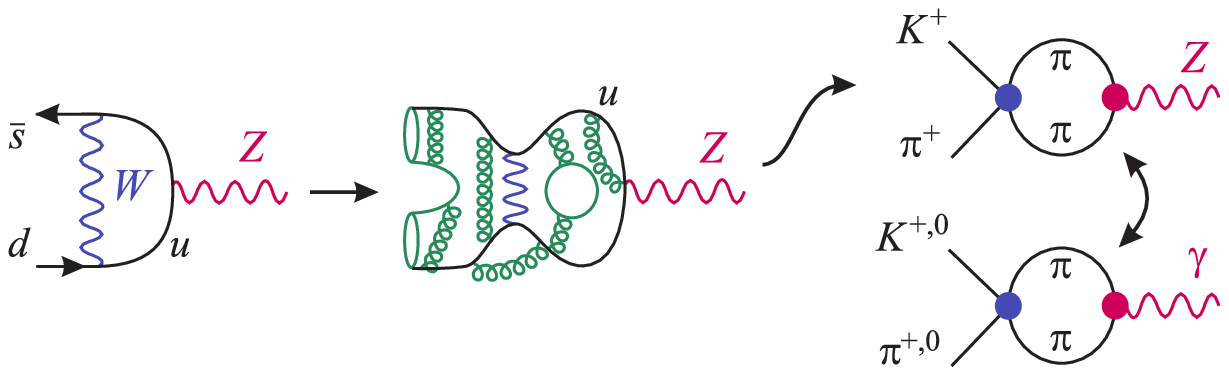}
\caption{The hadronic effects in the matrix elements of the short-distance top and charm-quark penguin operators are brought under control thanks to their relation with the charged-current-induced semileptonic $K_{\ell 3}$ processes~\cite{MesciaS07}. For the purely long-distance up-quark contribution to the Z penguin, the strategy relies on the photon penguin~\cite{IsidoriMS05}, which is entirely dominated by a similar up-quark contribution when CP-conserving (as in $K^+\rightarrow\pi^+\ell^+\ell^-$ and $K_S\rightarrow\pi^0\ell^+\ell^-$~\cite{DEIP98,BDI03}).}
\label{FigHadEff}
\end{figure}

\begin{itemize}
\item \textbf{Short-distance penguins and semileptonic decays}: Even though the charm and top quark penguins are effectively local point-like interactions at the kaon energy scale, they are still parametrized in terms of quark fields, whose hadronization into the initial kaon and final pion is not necessarily local. Technically, this step boils down to the evaluation of the hadronic matrix elements $\langle\pi^{0}|\bar{s}\gamma^{\mu}d|K^{0}\rangle$ and $\langle\pi^{+}|\bar{s}\gamma^{\mu}d|K^{+}\rangle$. The chiral symmetry relates these matrix elements to those relevant for the charged-current $K\rightarrow\pi\ell\nu$ decays, $\langle\pi^{+}|\bar{s}\gamma^{\mu}d|K^{0}\rangle$ and $\langle\pi^{0}|\bar{s}\gamma^{\mu}d|K^{+}\rangle$, see Fig.~\ref{FigHadEff}. Further, the impact of isospin symmetry breaking as well as of long-distance QED effects can be treated perturbatively. The precision achieved is such that these sources of hadronic uncertainties are very small, at the few percent-level for the $K\rightarrow\pi\nu\bar{\nu}$ decay rates~\cite{MesciaS07}. For comparison, remember that the lack of a similar strategy for the uncertainties on the matrix elements like $\langle\pi\pi|\bar{s}\Gamma d|K\rangle$~\cite{BGB14} is the main reason why the theoretical control over the $\varepsilon^{\prime}$ observable remains so challenging to this day.

\item \textbf{Long-distance penguins and radiative decays}: The second type of long-distance effects are the up-quark contributions to the penguins. Those are purely non-local, and have to be dealt with entirely in terms of meson external states and loops in the context of chiral perturbation theory. Precision is then limited by the rather slow convergence of the chiral expansions (around 30\% per order), and by the regular occurrence of free parameters, the counterterms, whose presence is often required to absorb loop divergences. However, as said before, these issues can be at least partly circumvented when, instead of calculating directly a given process from the chiral Lagrangian at a given order, one tries to relate it to well-measured observables. For example, the indirect CP-violating contribution $K_{L}\rightarrow\varepsilon(K_{1}\rightarrow\pi^{0}\ell^{+}\ell^{-})$ can be controlled~\cite{DEIP98,BDI03,FGD04} thanks to the measured $K_{S}\approx K_{1}\rightarrow\pi^{0}\ell^{+}\ell^{-}$ decay rates~\cite{NA48as}. Actually, this process also helps evaluating the up-quark contribution to $K^{+}\rightarrow\pi^{+}\nu\bar{\nu}$~\cite{IsidoriMS05}, see Fig.~\ref{FigHadEff}. Similarly, the experimental $K_{L}\rightarrow\pi^{0}\gamma\gamma$ rate and photon energy spectrum~\cite{KLpgg} are instrumental in controlling the two-photon contributions to $K_{L}\rightarrow\pi^{0}\ell^{+}\ell^{-}$~\cite{BDI03,ISU04}.
\end{itemize}

The current predictions for the rare decay rates are, in units of $10^{-11}$,%
\begin{equation}%
\begin{array}[c]{rllll}%
\mathcal{B}(K^{+}\overset{}{\rightarrow}\pi^{+}\nu\bar{\nu}(\gamma))^{\mathrm{SM}} & =8.25(64) & 
\text{\cite{BGS11}} & 
\;(\mathcal{B}^{\exp}=17.3_{-10.5}^{+11.5} & 
\text{\cite{KPpnunu}})\;\;,
\smallskip\\
\mathcal{B}(K_{L}\overset{}{\rightarrow}\pi^{0}\nu\bar{\nu})^{\mathrm{SM}} & =2.60(37) & 
\text{\cite{BGS11}} & 
\;(\mathcal{B}^{\exp}<2600 & \text{\cite{K0pnunu}})\;\;,
\smallskip\\
\mathcal{B}(K_{L}\overset{}{\rightarrow}\pi^{0}e^{+}e^{-})^{\mathrm{SM}} & =3.23_{-0.79}^{+0.91} & 
\text{\cite{MertensS11}\ \ \ \ \ } & 
\;(\mathcal{B}^{\exp}<28 & 
\text{\cite{KTeVelec}})\;,
\smallskip\\
\mathcal{B}(K_{L}\overset{}{\rightarrow}\pi^{0}\mu^{+}\mu^{-})^{\mathrm{SM}} & =1.29(24) & 
\text{\cite{MertensS11}} & 
\;(\mathcal{B}^{\exp}<38 & 
\text{\cite{KTeVmuon}})\;,
\end{array}
\end{equation}
The errors for the neutrino modes are dominated by the parametric uncertainties on the CKM matrix elements, which account for 59\% (82\%) of the total error for the $K^{+}$ ($K_{L}$) modes, respectively (see Ref.~\cite{BGS11} for a detailed breakdown of the errors into their various sources). On the other hand, the error for $\mathcal{B}(K_{L}\rightarrow\pi^{0}\ell^{+}\ell^{-})$ are dominated by that on the experimental $K_{S}\rightarrow\pi^{0}\ell^{+}\ell^{-}$ decay rates~\cite{ISU04,MertensS11}, with a smaller (and reducible with appropriate lepton momentum cuts) uncertainty coming from $K_{L}\rightarrow\pi^{0}\gamma\gamma$ for the muon mode.

The first message of this section is thus that the very clean $K\rightarrow\pi\nu\bar{\nu}$ decays do not suffer from large hadronic uncertainties. Currently, those should not prevent the observability of less than 10\% deviations from the SM predictions. At the same time, it should be clear that these modes are sensitive only to a limited class of NP effects since neutrinos interact only weakly in the SM. So, having at hand also other observables would be much welcome. For example, the $K_{L}\rightarrow\pi^{0}e^{+}e^{-}$ is very sensitive to NP deviations in the electromagnetic currents (both electric and magnetic) while the $K_{L}\rightarrow\pi^{0}\mu^{+}\mu^{-}$ would be sensitive to a new helicity-suppressed scalar current~\cite{MesciaST}. The second important message of this section is thus that for many modes, the strategies have been set to further improve their theoretical control using experimental inputs from other observables, especially the $K_{\ell3}$ decays for matrix elements~\cite{MesciaS07} and the radiative decays for the purely long-distance contributions. As discussed briefly above, this is the case for the $K_{L}\rightarrow\pi^{0}\ell^{+}\ell^{-}$ modes, but also to some extent for other processes like for example the $K_{L}\rightarrow\ell^{+}\ell^{-}$ rate~\cite{DAmbrosioIP98,DAmbrosioP97,IsidoriU04,GerardST05} or the $K^{+}\rightarrow\pi^{+}\pi^{0}\gamma$ CP-asymmetries~\cite{MertensS11,CappielloCDG12}. It is thus important to include in future experimental programs aiming at the rare decays at least also all the leading radiative modes $K_{S}\rightarrow\pi^{0}\gamma^{(\ast)}$, $K^{+}\rightarrow\pi^{+}\gamma^{(\ast)}$, $K_{L,S}\rightarrow(\pi^{0})\gamma^{(\ast)}\gamma^{(\ast)}$, $K^{+}\rightarrow\pi^{+}\gamma^{(\ast)}\gamma^{(\ast)}$.

\section{Supersymmetric effects in rare $K$ decays}

One of the most appealing classes of NP scenarios is based on supersymmetry. The most simple implementation, the Minimal Supersymmetric Standard Model (MSSM), is actively looked for at the LHC. Currently, the bounds on the gluino and squark masses, assuming they are all degenerate, are in the $1-2$ TeV range. This is rather high from the point of view of low-energy flavor observables. Taken at face value, and without any large departure from MFV, the impact on the rare $K$ decays would be far too small to be observed.

In the following, we will assume that MFV is at least approximately valid. Then, the only way supersymmetry could show up in rare $K$ decays is if at least some of the sparticles are below the TeV scale, hence below the naive mass bounds set by the LHC. This is possible only in special circumstances, two of which will be briefly discussed here.

\subsection{Natural SUSY and Minimal Flavor Violation\label{NsusyMFV}}

In the MSSM, besides the fact that two Higgs doublets are needed, the spontaneous breaking of the electroweak symmetry becomes supersymmetric in the sense that it is driven by the parameters of the supersymmetric Lagrangian. Further, it cannot occur without breaking first the supersymmetry itself. As a result, the conditions for a consistent electroweak symmetry breaking entangle the various scales at play, schematically as
\begin{equation}
m_{Z}^{2}\sim\mu^{2}+m_{H_{u}}^{2}\;, \label{EWscale}
\end{equation}
where $m_{Z}$ is the $Z$ boson mass (electroweak scale), $\mu$ the supersymmetric Higgs mass term (SUSY scale), and $m_{H_{u}}$ the soft supersymmetry breaking Higgs mass term (SUSY breaking scale). The last two parameters also drive the mass of the sparticles, so should a priori not be too small. At the same time, they must be adjusted so as to maintain the $Z$ boson mass at the much lower electroweak scale. This fine tuning quickly becomes horrendous as $\mu$ and $m_{H_{u}}$ increase. Actually, even the parameters tuning the loop corrections to these mass terms have to be controlled.

In the so-called Natural SUSY scenario (see e.g. Ref.~\cite{NSUSY} and references therein), one insists that the electroweak symmetry breaking conditions remain natural, i.e., free from a too serious fine-tuning. The SUSY spectrum must then satisfy some generic conditions. In particular, the top squarks must not be too heavy. This is compatible with the LHC constraints because the mass bounds hold mostly for the light flavor squarks, which could be more directly produced out of the first-generation quarks in the proton. The gluino must also be rather light, not beyond about $1.5$ TeV. When combined with a naive unification conditions for gaugino masses, this further implies the presence of at least one rather light neutralino and chargino.

This is not the full story yet. The supersymmetrization of the electroweak symmetry breaking has another important consequence: it constrains the Higgs boson mass which has to be below $m_{Z}$ at tree level. Here, somewhat contrary to what is needed to avoid a too large fine-tuning in Eq.~(\ref{EWscale}), a large loop correction is required to reproduce the observed mass of about $125$ GeV. The main contribution comes from the top sector:
\begin{equation}
m_{h}^{2}=m_{Z}^{2}\cos2\beta+\frac{m_{t}^{4}}{v_{\mathrm{EW}}^{2}}\Delta_{loop}^{t,\tilde{t}}\;,\; \label{HiggsMass}
\end{equation}
with
\begin{equation}
\Delta_{loop}^{t,\tilde{t}}\sim a\log\frac{m_{\tilde{t}_{1,2}}}{m_{t}}+b\frac{X_{\tilde{t}_{L}-\tilde{t}_{R}}^{2}}{m_{\tilde{t}_{1,2}}^{2}}\left(1-\frac{X_{\tilde{t}_{L}-\tilde{t}_{R}}^{2}}{12m_{\tilde{t}_{1,2}}^{2}}\right)  \;,
\end{equation}
where $a$ and $b$ are some coefficients, $m_{\tilde{t}_{1,2}}$ the two stop masses, and $X_{\tilde{t}_{L}-\tilde{t}_{R}}$ their mixing. So, the stop should thus not be too light, and this mixing should better be rather large.

\begin{figure}[t]
\centering                 \includegraphics[width=15cm]{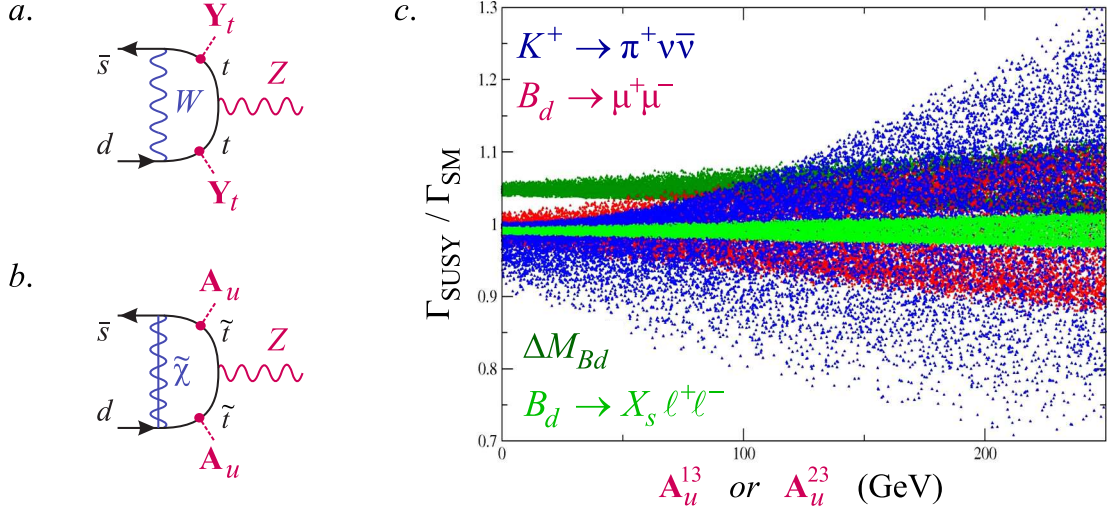}
\caption{The SM top quark contribution to the $Z$ penguin $(a.)$ and its
chargino-stop supersymmetric analogue $(b.)$. The flavor transition is driven
by a double Yukawa insertion (quadratic GIM breaking) in the SM, and by a
double up-type trilinear insertion in the SUSY case. As shown on the right
$(c.)$, the $K\rightarrow\pi\nu\bar{\nu}$ modes are then particularly sensitive
to the trilinear matrix elements (see Ref.~\cite{IMPST06} for more details).}%
\label{SUSYpeng}%
\end{figure}

At the end of the day, one thus remains with rather light stop and gauginos, and a possibly significant up-type trilinear term $\mathbf{A}_{u}$ driving the stop mixing through $(\mathbf{A}_{u}^{\dagger})_{33}$. Such a situation is ideal to generate a large deviation in the $Z$ penguin as the supersymmetric analogue of the SM contribution is enhanced, see Fig.~\ref{SUSYpeng}. Indeed, this correction is driven by chargino-stop loop~\cite{Chargino,ColangeloI}, which would both be light, and by a double insertion $(\mathbf{A}_{u})_{13}(\mathbf{A}_{u}^{\dagger})_{32}$, which could be sizable~\cite{ColangeloI}. Actually, the $K\rightarrow\pi\nu\bar{\nu}$ modes are the best probes for such a deviation~\cite{IMPST06}, as shown in Fig.~\ref{SUSYpeng}. There is however a caveat to keep in mind. If MFV is active, then $\mathbf{A}_{u}\approx A_{0}\mathbf{Y}_{u}$ with $A_{0}$ setting the SUSY breaking scale. The double trilinear insertion is again tuned by the CKM matrix, $(\mathbf{A}_{u})_{13}(\mathbf{A}_{u}^{\dagger})_{32}\sim V_{td}V_{ts}^{\dagger}$, and is thus suppressed. In those instance, the SUSY\ correction to the $Z$ penguin is limited, even for relatively light stops and charginos~\cite{MFVapp,IMPST06}.

The question at this stage is thus whether MFV is compatible with a natural SUSY like scenario, where the squark masses are highly split. This issue was analyzed recently in Ref.~\cite{NSUSYMFV}, with the following conclusions:

\begin{itemize}
\item As a preliminary, it is important to realize that a split squark spectrum is not so trivial to parametrize. Which scalar quark states are to be called stops depends on the form of the Yukawa couplings in the gauge flavor basis, which we cannot access. In other words, all the flavor couplings (Yukawa couplings and soft supersymmetry breaking terms) should in principle be known in a given basis to unambiguously fix the squark spectrum. Short of a flavor model, it is however always possible to circumvent this difficulty by writing the soft-supersymmetry breaking terms as polynomial expansions in the Yukawa couplings~\cite{MFVNiko}. Doing this ensures that they depend on the flavor basis in a coherent and consistent way~\cite{NSUSYMFV}.

\item From a phenomenological point of view, a split third generation of squarks is easily parametrized. For example, light left stop and sbottom could result from
\begin{equation}
\text{Split }\tilde{t}_{L},\tilde{b}_{L}:\mathbf{m}_{Q}^{2}=m_{0}^{2}(\mathbf{1}-\mathbf{Y}_{u}^{\dagger}\mathbf{Y}_{u}/\langle\mathbf{Y}_{u}^{\dagger}\mathbf{Y}_{u}\rangle)\;, 
\label{Nat1}
\end{equation}
where $\mathbf{m}_{Q}^{2}$ is the left-squark soft mass term, $\langle...\rangle$ denotes the matrix trace, and $m_{0}$ sets the SUSY breaking scale. In the up-quark mass eigenstate basis, $\mathbf{Y}_{u}$ is diagonal, so $m_{Q}^{2}\approx m_{0}^{2}(1,1,0)$ in that basis. The same strategy can be used to split the right stop or sbottom with e.g.
\begin{align}
\text{Split }\tilde{t}_{R}  &  :\mathbf{m}_{U}^{2}=m_{0}^{2}(\mathbf{1}-\mathbf{Y}_{u}\mathbf{Y}_{u}^{\dagger}/\langle\mathbf{Y}_{u}\mathbf{Y}_{u}^{\dagger}\rangle)\;,\label{Nat2}\\
\text{Split }\tilde{b}_{R}  &  :\mathbf{m}_{D}^{2}=m_{0}^{2}(\mathbf{1}-\mathbf{Y}_{d}\mathbf{Y}_{d}^{\dagger}/\langle\mathbf{Y}_{d}\mathbf{Y}_{d}^{\dagger}\rangle)\;. 
\label{Nat3}
\end{align}
Note that the form of these polynomials are dictated by the transformation properties of the flavor couplings under the $SU(3)^{5}$ flavor symmetry~\cite{Georgi} of the gauge interactions.

\item Such polynomial expansions respect the MFV hypothesis when all coefficients are $\mathcal{O}(1)$ numbers~\cite{MFVMercolli}. So, for reasonable values of $\tan\beta$, a split $\tilde{t}_{R}$ is necessarily MFV-like since $\langle\mathbf{Y}_{u}^{\dagger}\mathbf{Y}_{u}\rangle$ is such an $\mathcal{O}(1)$ number. On the other hand, split $\tilde{b}_{R}$ is compatible with MFV only for very large $\tan\beta$, since otherwise $\langle\mathbf{Y}_{d}\mathbf{Y}_{d}^{\dagger}\rangle^{-1}\gg\mathcal{O}(1)$. Finally, split $\tilde{t}_{L},\tilde{b}_{L}$ may or may not respect MFV, depending on $\tan\beta$ and whether they are split using a $\mathbf{Y}_{u}^{\dagger}\mathbf{Y}_{u}/\langle\mathbf{Y}_{u}^{\dagger}\mathbf{Y}_{u}\rangle$ term or a $\mathbf{Y}_{d}^{\dagger}\mathbf{Y}_{d}/\langle\mathbf{Y}_{d}^{\dagger}\mathbf{Y}_{d}\rangle$ term (both are permitted by the flavor symmetry).

\item If the squark splitting is induced by boundary conditions on the soft terms at the GUT scale, then a non-MFV structure in one soft-term can pollute the others through the running. On the other hand, generically, any non-MFV structure tends to be suppressed running down~\cite{MFVrunning}. As a result, even a scenario like Eq.~(\ref{Nat3}) at the GUT scale could be reasonably compatible with MFV at the low scale.
\end{itemize}

In conclusion, the rare $K\rightarrow\pi\nu\bar{\nu}$ modes are ideally suited to probe natural SUSY like scenarios where charginos and stops could be relatively light. On the one hand, they are the most sensitive to the up-type squark mixing terms, and on the other, their exceptional cleanness put them in the best situation to discern the small deviations expected whenever MFV\ is active.

\subsection{Baryon number violating SUSY}

Once MFV comes into play in the MSSM, there are two notable consequences on the phenomenology. First, the supersymmetric corrections to the rare $K$ decay rates are suppressed by the small CKM elements. In addition, they quickly decouple and would be totally unobservable for sparticles around the TeV. To some extent, this remains true for flavor physics in general: under MFV, the sparticles circulating in an FCNC loop should better be well below the TeV scale if they are to induce a noticeable shift with respect to the SM. Somehow, these sparticles must thus have escaped detection at the LHC.

Coincidentally, the second consequence of MFV offers precisely such a hiding mechanism. Phenomenologically, MFV is redundant with R parity~\cite{MFVRPV} in the sense that it suffices to prevent a too fast proton decay (or neutron oscillation). So, once MFV is brought in, the main incentive to impose this ad-hoc discrete symmetry disappears. But once R parity is no longer imposed~\cite{Barbier04}, all the sparticles ultimately decay into SM particles, and the telltale missing energy channels used to look for supersymmetry at colliders are no longer adequate~\cite{MFVRPVCGH}.

The hiding power of MFV is even stronger. From a flavor symmetry point of view, the R-parity couplings violating baryon ($\mathcal{B}$) and lepton ($\mathcal{L}$) number have intrinsically different characteristics. The best way to see this is to first answer the following~\cite{MFV4G}: is it possible to construct flavor blind couplings which break lepton and/or baryon number? Such couplings should be invariant under $SU(3)^{5}$, but not under the full $U(3)^{5}$ flavor symmetry of the gauge interactions~\cite{Georgi}, since the $U(1)$s of $\mathcal{B}$ and $\mathcal{L}$ are combinations of $U(1)$ transformations acting on the individual quark and lepton fields. A flavor-blind coupling is thus a contraction of quark and lepton fields with the invariant tensors of the $SU(3)$ groups. Clearly, the kronecker delta never leads to a $\Delta\mathcal{B}$ or $\Delta\mathcal{L}$ coupling since it is also an invariant of $U(3)^{5}$. In practice, these contractions are quark-antiquark or lepton-antilepton pairs, without any $\mathcal{B}$ or $\mathcal{L}$ charge. On the other hand, the Levi-Civita tensor contracts three quark or lepton fields together, hence does have a total $\mathcal{B}$ or $\mathcal{L}$ charge of three units\footnote{For historical reasons, the elementary $\mathcal{B}$ unit is $1/3$, so that nucleons have $\mathcal{B}=1$. The elementary $\mathcal{L}$  unit is that of the electron, $\mathcal{L}=1$.}. The best known example of such a flavor blind interaction is the $\mathcal{B}+\mathcal{L}$ anomaly of the SM~\cite{tHooft76}, which corresponds to four such flavor-blind contractions among left-handed quark $Q$ and lepton doublets $L$~\cite{MFV4G}:
\begin{equation}
H_{eff}\sim(\varepsilon^{IJK}Q^{I}Q^{J}Q^{K})\times(\varepsilon^{IJK}Q^{I}Q^{J}Q^{K})\times(\varepsilon^{IJK}Q^{I}Q^{J}Q^{K})\times(\varepsilon^{IJK}L^{I}L^{J}L^{K})\;,
\end{equation}
where $I,J,K=1,2,3$ are flavor indices.

Returning to the MSSM, none of the R-parity violating coupling is flavor-blind, since they all involve several fermion species transforming under different $SU(3)$s. But, once Yukawa interactions are allowed, Yukawa couplings, transforming under several $SU(3)$s, can be introduced to construct a flavor-symmetric coupling. This is possible only if the underlying coupling breaks $\mathcal{B}$ or $\mathcal{L}$ by three units. The R-parity violating couplings breaking $\mathcal{L}$ by one unit\footnote{In the presence of a $\Delta\mathcal{L}=2$ Majorana mass term, the $\Delta\mathcal{L}=1$ couplings are allowed but tiny since they end up tuned by the left-handed neutrino masses.} are thus forbidden, and proton decay does not occur. Remains only the $\Delta \mathcal{B}=1$ coupling $\mathbf{Y}_{udd}^{IJK}U^{I}D^{J}D^{K}$, for which $SU(3)^{5}$ symmetric representations can be constructed as for example~\cite{MFVRPV,MFVRPVrun}
\begin{subequations}
\begin{align}
(\mathbf{Y}_{udd})^{IJK}  &  \sim\varepsilon^{LMN}\mathbf{Y}_{u}^{IL}\mathbf{Y}_{d}^{JM}\mathbf{Y}_{d}^{KN}+...\;,\label{BasicQ}\\
(\mathbf{Y}_{udd})^{IJK}  &  \sim\varepsilon^{LJK}(\mathbf{Y}_{u}\mathbf{Y}_{d}^{\dagger})^{IL}+...\;. \label{BasicU}
\end{align}
Phenomenologically, the underlying epsilon contraction combined with the quasi-diagonal CKM matrix forces these couplings to be largest when the three quarks are of three different generations. Actually, the largest couplings all involve a top flavored quark or squark. Hence, the largest couplings do not contribute to low-energy $\Delta\mathcal{B}$ observables like neutron oscillations.

\begin{figure}[t]
\centering                \includegraphics[width=15cm]{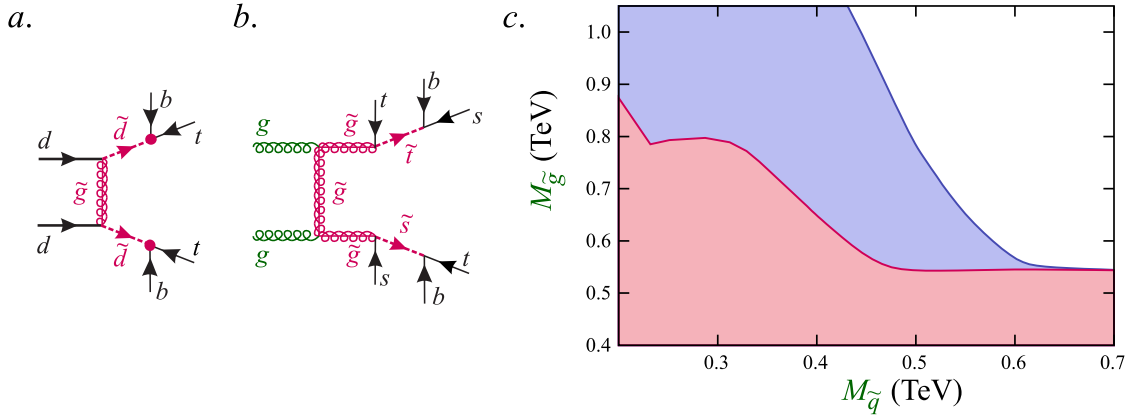}
\caption{Some QCD production mechanisms for squarks $(a.)$ and gluinos $(b.)$,
and their subsequent decay through the dominant MFV R-parity violating
couplings. $c.)$ Corresponding mass bounds on the first generation squarks and
gluinos derived from the generic NP searches in the same-sign lepton pair
channels by the CMS collaboration. The blue and red exclusion areas denote
different mass spectrum (hence decay chains) hypotheses. See
Ref.~\cite{MFVLHC2} for more details.}
\label{FigRPV}
\end{figure}

At colliders, all sparticle decay chains end with an intense hadronic activity instead of the standard missing energy signatures. This looks like a particularly effective way to hide supersymmetry at hadron machines, but for one special window. The largest couplings involve the top flavor. So, pairs of same-sign top quarks are easily produced through channels like $gg\rightarrow \tilde{g}\tilde{g}\rightarrow tt+$jets or $dd\rightarrow\tilde{d}\tilde{d}\rightarrow\bar{t}\bar{t}+$jets, see Fig.~\ref{FigRPV}. In turn, such top quark pairs produce same-sign lepton pairs, a signature scrutinized at the LHC~\cite{MFVLHC1,MFVLHC2}. As a result, there are already some mass bounds on these scenarios, see Fig.~\ref{FigRPV}, but those are indeed far below those set in the R-parity conserving case. 

Such a pattern of RPV leaves more room for sizable effects in flavor physics since sparticles can be lighter. Apart from that, the supersymmetric phenomenology is not much affected. Indeed, the $\mathbf{Y}_{udd}$ couplings are rather suppressed, especially for Eq.~(\ref{BasicQ}), and their impact on the rest of the MSSM as well as on the FCNC is quadratic. For example, it is trivial to render the natural SUSY setting discussed before R-parity violating. Neither the fine-tuning in Eq.~(\ref{EWscale}) nor the Higgs mass Eq.~(\ref{HiggsMass}) are significantly affected. But, now that both the chargino and stop can be lighter, their contribution to the rare $K$ decays can be larger, improving the prospects for an experimental discovery.

Before closing this summary of the baryon number violating MSSM, let us comment on one last feature of this scenario. If for some reason holomorphy is assumed to hold also for the Yukawa coupling insertions, then there is only one way to parametrize the RPV coupling: $(\mathbf{Y}_{udd})^{IJK} \sim\varepsilon^{LMN}\mathbf{Y}_{u}^{IL}\mathbf{Y}_{d}^{JM}\mathbf{Y}_{d}^{KN}$~\cite{MFVRPVCGH}. All the other $SU(3)^5$-symmetric constructions involve either $\mathbf{Y}_{u}^{\dagger}$ or $\mathbf{Y}_{d}^{\dagger}$, hence would not be ``spurion-holomorphic''. Recently, this scenario was found to have a very intriguing property~\cite{MFVRPVrun}: it retains its form under the renormalization group evolution, to all orders. In other words, we can write simply
\end{subequations}
\begin{equation}
(\mathbf{Y}_{udd}[Q])^{IJK}=\lambda\lbrack Q]\varepsilon^{LMN}\mathbf{Y}_{u}[Q]^{IL}\mathbf{Y}_{d}[Q]^{JM}\mathbf{Y}_{d}[Q]^{KN}\;.
\end{equation}
The whole renormalization group evolution of $\mathbf{Y}_{udd}$ is encoded in that of the Yukawa couplings and in its coefficient $\lambda$. Contrary to all the other ways to construct a flavor-symmetric $UDD$ coupling, this is the only one for which additional terms never arise\footnote{Specifically, corrections like $\varepsilon^{LMN}(\mathbf{Y}_{u}\mathbf{Y}_{u}^{\dagger}\mathbf{Y}_{u})^{IL}\mathbf{Y}_{d}^{JM}\mathbf{Y}_{d}^{KN}$, $\varepsilon^{LMN}\mathbf{Y}_{u}^{IL}(\mathbf{Y}_{d}\mathbf{Y}_{u}^{\dagger}\mathbf{Y}_{u})^{JM}\mathbf{Y}_{d}^{KN}$, etc, do arise, but they all either compensate each other, sum up to reproduce the evolution of the Yukawa couplings, or combine into a flavor-blind matrix trace absorbed into the coefficient $\lambda$.}. Once imposed, holomorphy is preserved even though the Yukawa couplings are not dynamical fields but mere spurions. Further, if non-holomorphic terms like for example $\varepsilon^{LMN}(\mathbf{Y}_{u}\mathbf{Y}_{u}^{\dagger}\mathbf{Y}_{u})^{IL}\mathbf{Y}_{d}^{JM}\mathbf{Y}_{d}^{KN}$ were present at some high scale, they would be washed away through the running, and effectively only the holomorphic structure would be relevant at the low scale. In this particular instance, the purely phenomenological MFV hypothesis has given birth to a truly dynamical principle for the R-parity violating MSSM.

\section{Rare $K$ decays and dark portals\label{portals}}

Since the SM is so succesful, most scenarios of NP are introduced as alternative UV completions. The dynamics above (but not too far from) the electroweak scale is altered by the presence of new interactions or forms of matter. But at or below the electroweak scale, all the new states decouple and their effects can be fully encoded into new interactions among SM particles only. The leading corrections are parametrized by dimension-six effective operators~\cite{BuchmullerW86}, hence are relatively suppressed\footnote{If $\mathcal{L}$ is broken, there is an additional dimension-five operator leading to a Majorana mass term for the neutrinos\cite{Weinberg}.}.

Strictly speaking though, the successes of the SM do not rule out the presence of additional states at low energy. Provided they interact very weakly, their impact would be minimal, and they could have escaped detection up to now. Actually, many scenarios predict the existence of such new light states. Let us mention for example the axions, familons, dark photons, millicharged fermions, dilaton, majoron, sterile neutrino, gravitino,... (see e.g. Ref.~\cite{JaeckelR10} for a review). In addition, if one or more of these states is sufficiently stable, it could contribute to the dark matter density. 

Experimentally, to discover such new states is tricky. They would show up as missing energy $\slashed E$, and would thus be undistinguishable from neutrinos but for the different kinematics when massive. In addition, being weakly interacting, high luminosity is crucial and except in some special circumstances, colliders cannot compete with low-energy experiments yet. So, with the advent of a new generation of experiments aiming at the very rare $K\rightarrow\pi\nu\bar{\nu}$ modes, it is timely to reconsider the constraints one could draw on the presence of new light states. Besides,  it is worth to remember that historically, rare $K$ decays have already played a central role in this context: the non-observation of $K\rightarrow\pi+\slashed  E$ ruled out the simplest axion model~\cite{WeinbergA}.

\begin{figure}[t]
\centering                \includegraphics[width=11cm]{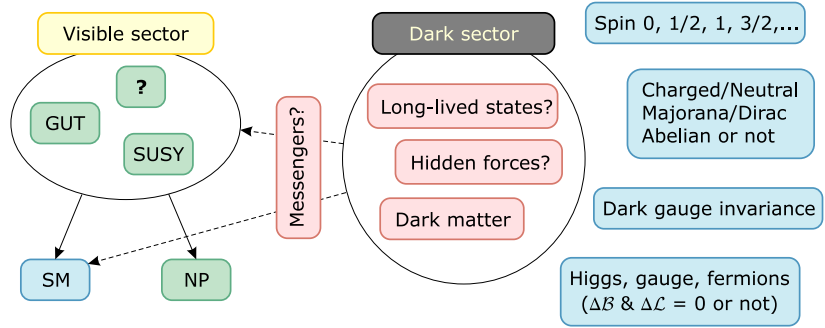}
\caption{Schematic representation of a dark sector. Besides the corrections to the SM induced by the ``visible'' new physics, parametrized at low-energy through higher-dimensional operators among SM fields~\cite{BuchmullerW86}, the presence at low energy of a new state, whether it is a messenger, a long-lived particle, or a dark matter constituent, has to be parametrized though new operators whose leading mass-dimension strongly depend on their nature and on generic hypotheses about their interactions with the SM.}%
\label{Dark}%
\end{figure}

To organize the search for such light states as model-independently as possible, a possible strategy is to construct the equivalent of the Buchmuller-Wyler operator basis~\cite{BuchmullerW86} once the SM particle content is extended, and then constrain all the operators involving the new state(s), see Fig.~\ref{Dark}. This program is more involved than it seems for several reasons. First, the leading operators to consider, the so-called portals, strongly depend on generic assumptions which have to be made on the nature of the new state. Evidently, its spin has to be specified, as well as whether it carries a dark charge and needs to be pair produced. For vector (or higher spin states), the presence of an underlying dark gauge invariance strongly constrains the form of the leading operators.

From the point of view of flavor physics, another important issue is whether these new states couple dominantly to Higgs or gauge bosons, hence are flavor-blind, or when they couple to quarks and leptons, whether they are able to directly induce the flavor transition. Specifically, we can consider three classes of scenarios for a generic effective coupling of the dark state(s) $X$ to quarks
\[
\begin{tabular}
[c]{ccc}\hline
Flavor-changing & \multicolumn{2}{c}{Flavor-blind $\{q^{I}\Gamma q^{I}\}\otimes X\,$}\\\cline{2-3}
$\{q^{J}\Gamma q^{I}\}\otimes X\,,\;I\neq J\,$ & Heavy quarks & Light quarks\\\hline
\raisebox{-0.4108in}{\includegraphics[height=0.8804in,width=1.1519in]{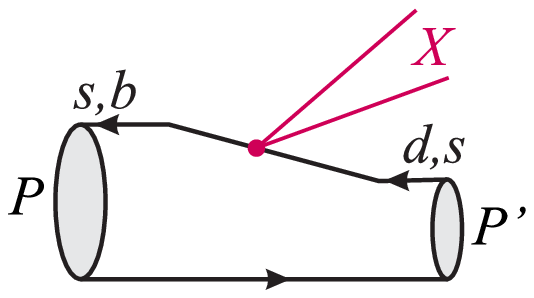}}
$\;\;\;\;\;$ & \ \ \ \ \
\raisebox{-0.3918in}{\includegraphics[height=0.8795in,width=1.1519in]{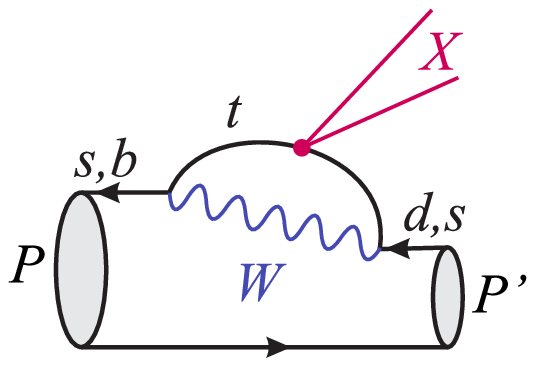}}
$\;\;$ & $%
\raisebox{-0.3918in}{\includegraphics[height=0.8795in,width=1.1597in]{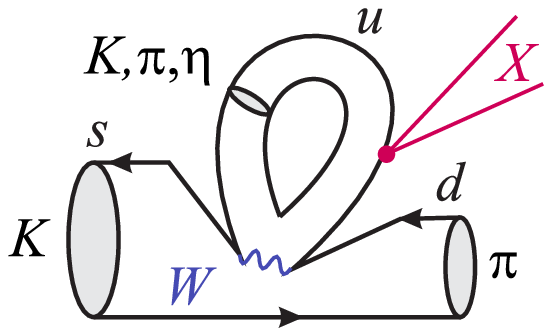}}
\;\;\;$\\
\multicolumn{2}{c}{Best probed through$\;\;\;$} & Sometimes competitive with\\
\multicolumn{2}{c}{FCNC observables$\;\;\;$} & flavor-blind searches\\
&  & (quarkonia, EWPO, beam dump,...)\\\hline
\end{tabular}
\ \ \ \ \
\]
In Ref.~\cite{KamenikS11}, the full set of operators for the various assumptions on the dark state $X$ were constructed, and the constraints drawn from $B$ and $K$ decays with missing energy in the final states compared. We do not intend to repeat this analysis here, but instead illustrate two particular aspects that need to be kept in mind in designing experimental search strategies.

\subsection{Flavor-breaking portals: Beware of the kinematics}

\begin{figure}[t]
\centering      \includegraphics[width=15cm]{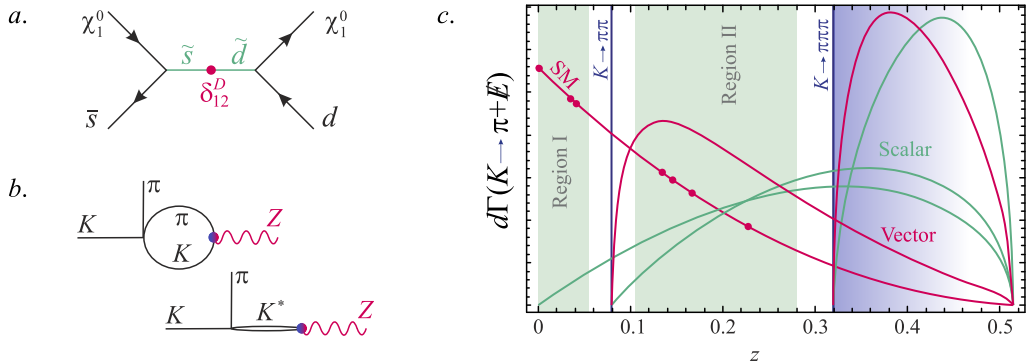}  \caption{$a.)$
Tree-level pair production of neutralinos, with the flavor transitions induced
by the squark soft-supersymmetry breaking terms. $b.)$ The $K\pi$ meson loop
driving the momentum dependence of the form-factor for the SM vector current.
In the $p$-wave, the $K\pi$ interaction is strongly influenced by vector
mesons, whose lowest lying state is the $K^{\ast}(892)$. $c.)$ The differential
spectra as a function of $z=(p_{\nu}+p_{\bar{\nu}})^{2}/m_{K}^{2}$, see
Eq.~(\ref{KinKpnn}). The red and green curves correspond to the vector and
scalar currents of Eq.~(\ref{HidVector}) and~(\ref{HidScalar}), respectively,
plotted for $M_{\chi}=m_{\pi},m_{\pi}/2,$ and $0$. The massless vector case
coincides with the SM, on which curve the E787 and E949 events are
reported~\cite{KPpnunu}. Region I and II are the planned windows of
observation of the NA62 experiment.}
\label{Kkin}
\end{figure}

As an example of new flavor-changing invisible decay channels, let us consider the very light neutralino scenario of Ref.~\cite{Neutralinos}. In the MSSM with R parity, the lightest supersymmetric particle is perfectly stable, and thus should be neutral for cosmological reasons. In many cases, the lightest of the four neutralinos could play this role. If light enough, the $K\rightarrow\pi\nu\bar{\nu}$ decay process would be accompanied by $K\rightarrow\pi\chi_{1}^{0}\chi_{1}^{0}$, since these neutralinos would also show up as missing energy. In the MSSM, the $K\rightarrow\pi\chi_{1}^{0}\chi_{1}^{0}$ process can arise at tree-level, see Fig.~\ref{Kkin}. Except for massless neutralinos, two types of effective interactions are generated: a scalar and a vector current,
\begin{align}
\text{Vector}  &  :\bar{s}\gamma_{\mu}(1\pm\gamma_{5})d\otimes\bar{\chi}^0_1\gamma_{\mu}\gamma_{5}\chi^0_1\;,\;\;\text{tuned by }\delta_{LL}^{sd},\delta_{RR}^{sd}\;,\label{HidVector}\\
\text{Scalar}  &  :\bar{s}(1\pm\gamma_{5})d\otimes\bar{\chi}^0_1(1\pm\gamma_{5})\chi^0_1\;,\;\;\text{tuned by }\delta_{LR}^{sd}\;.
\label{HidScalar}
\end{align}
In both cases, the underlying flavor transition is not driven by the SM flavor structures, but directly by the squark flavor mixing arising from the soft supersymmetry breaking terms. Numerically, this scenario is rather contrived though: the neutralinos must be really light, at the cost of a fine-tuning of the MSSM parameters, squarks should not be far above the electroweak scale and their flavor mixing should strongly deviate from MFV to induce an observable deviation with respect to the SM, in apparent conflict with the generic LHC mass bounds and with other flavor observables like kaon mixing or even $B$ physics.

This scenario may not be the most likely candidate for the physics beyond the SM, but it is however perfectly suited to illustrate an important point. One should never blindly use the $K\rightarrow\pi\nu\bar{\nu}$ rate to constrain a $K\rightarrow\pi+\slashed E$ process. First, the experimentalists look for the $K\rightarrow\pi\nu\bar{\nu}$ events only in two kinematical regions in pion momenta where the tremendous background can be managed: below the $K\rightarrow\pi\pi\pi$ threshold and avoiding the $K\rightarrow\pi\pi$ peak, see Fig.~\ref{Kkin}. The neutralino mass must thus be compatible with these windows. Second, the SM differential rate is built in their analysis strategies, but does not necessarily match that of neutralino production, see Fig.~\ref{Kkin}. How to translate the experimental result on $K\rightarrow\pi\nu\bar{\nu}$ into a bound on the production of new invisible states is thus far from trivial.

At this stage, it is worth to recall how the $K\rightarrow\pi\nu\bar{\nu}$ differential rate for the SM shown in Fig.~\ref{Kkin} is predicted. It originates from the hadronic matrix elements for the $Z$ penguin effective operator,
\begin{equation}
\langle\pi^{+,0}(p_{\pi})|\bar{s}\gamma^{\mu}d|K^{+,0}(p_{K})\rangle
=f^{++,00}_{+}(z)(p_{K}+p_{\pi})^{\mu}+f^{++,00}_{-}(z)(p_{K}-p_{\pi})^{\mu}\;,
\end{equation}
with $z=(p_{\nu}+p_{\bar{\nu}})^{2}/m_{K}^{2}$, leading to
\begin{equation}
\frac{d}{dz}\ln\Gamma(K^{+,0}\rightarrow\pi^{+,0}\nu\bar{\nu})\sim\frac{|\mathbf{p}_{\pi
}|^{3}}{m_{K}^{3}}\frac{|f^{++,00}_{+}(z)|^{2}}{|f^{++,00}_{+}(0)|^{2}}\;,\;\;|\mathbf{p}_{\pi}|=\frac{m_{K}}{2}\lambda^{1/2}(1,z,m_{\pi}^{2}/m_{K}^{2})\;,\; 
\label{KinKpnn}
\end{equation}
and the standard kinematical function is $\lambda(a,b,c)=a^{2}+b^{2}+c^{2}-2(ab+ac+bc)$. The form-factors $f^{++,00}_{+}(z)$ entering in this expression encode the coupling of the vector current to mesons (see Fig.~\ref{Kkin}). Their slopes are not fixed by the chiral symmetry but have to be measured. As for the electromagnetic current, to which they are directly related in the $SU(3)$ limit, they turn out to be mainly driven by vector meson exchanges. So, to an excellent approximation, the charged and neutral $K\rightarrow\pi$ transition form-factors have the same slope, driven by the pole of the neutral $K^{\ast}(892)$. Numerically, an estimate is derived from the slopes measured in the $K_{\ell 3}$ decays (see Fig.~\ref{FigHadEff}). This is reliable because the isospin breaking effect is tiny, below the percent level~\cite{MesciaS07}, as can be guessed from the small mass difference between $K^{\ast0}(892)$ and $K^{\ast+}(892)$.

Being entangled in the experimental analysis, this SM spectrum is assumed when the total $K\rightarrow\pi+\slashed  E$ rate is extrapolated from the events found in the two observation regions. In case NP modifies the spectrum, and short of a true measurement of the differential rate, one possible way to evidence it would be to compare the extrapolated rates using only the events either below or above the $K\rightarrow\pi\pi$ peak. As can be guessed from Fig.~\ref{Kkin}, and with sufficient statistics, this asymmetry would already probe a large range of scenarios.

\subsection{Flavor-blind portals: Go radiative}

Let us now consider a flavor blind scenario. Imagine that a new light vector boson couples to the light $u$, $d$, and $s$ quarks. Being flavor-blind, the $d$ and $s$ charges are the same. Without loss of generality, the light quark dark charges can then be aligned with either their electric charges $Q$ or their baryon number $\mathcal{B}$~\cite{Fayet89}:
\begin{equation}
\mathcal{H}_{eff}=e 
\left( \begin{array}[c]{ccc} \bar{u} & \bar{d} & \bar{s} \end{array} \right) \slashed V 
\left(  
\frac{\varepsilon_{Q}}{3} \left( \begin{array}[c]{ccc} 2 &  & \\ & -1 & \\ &  & -1 \end{array} \right)  
+\varepsilon_{\mathcal{B}} \left(\begin{array}[c]{ccc} 1 &  & \\ & 1 & \\ &  & 1 \end{array} \right)  
\right)  
\left( \begin{array}[c]{c} u\\ d\\ s \end{array} \right)\,,
\label{DarkVC}
\end{equation}
with $e$ the positron electric charge. The goal is to extract bounds on $\varepsilon_{Q}$ and $\varepsilon_{\mathcal{B}}$ from $K$ decays, and check whether they are competitive with those arising from flavor-blind observables. For $\varepsilon_{Q}$, if the dark vector also couples to leptons, as they usually do when arising e.g. from a $U(1)$ kinetic mixing term, a bound below $10^{-3}$ should be aimed at (see e.g. Refs.~\cite{JaeckelR10,Essig,Williams} for reviews). The same is true for $\varepsilon_{\mathcal{B}}$ if these interactions respect a $\mathcal{B}-\mathcal{L}$ symmetry, see e.g. Ref.~\cite{Heeck}. Note, though, that from a purely model-independent perspective, dark vectors may couple predominantly to quarks, in which case $\varepsilon_{Q}$ and $\varepsilon_{\mathcal{B}}$ are not much constrained yet, especially for dark vectors of a few hundred MeV in mass. In that case, $K$ decays are in a good position to play the leading role.

The dark vector current Eq.~(\ref{DarkVC}) can directly be embedded within Chiral Perturbation Theory (ChPT). As discussed in Ref.~\cite{KamenikS11}, the situation is then very different for $\varepsilon_{Q}$ and $\varepsilon_{\mathcal{B}}$. The former is perfectly aligned with the photon current, hence the predicted rates for any production mechanism can directly be deduced from the corresponding rate with photon(s), up to phase-space corrections. Assuming the dark vector is light enough to neglect these phase-space differences,
\begin{equation}
\mathcal{B}(K\rightarrow a\pi+b\gamma+cV)
\approx\varepsilon_{Q}^{2c}\mathcal{B}(K\rightarrow a\pi+(b+c)\gamma)\;,
\end{equation}
for $a=1,2,3$, and $b,c$ positive integers. Since branching ratios decrease with increasing number of pions and photons, only the simplest modes are worth considering. To derive bounds on $\varepsilon_{\mathcal{B}}$ is far less trivial. The baryon-number vector current decouples from mesons at leading order in $SU(3)$ ChPT. At the next to leading order in the momentum expansion, it enters through unknown counterterms, and more interestingly, through the parity-odd anomalous interactions (alongside $\varepsilon_{Q}$). So, we will only consider parity-odd observables to constrain $\varepsilon_{\mathcal{B}}$, and more specifically, processes induced through $\pi^{0},\eta,\eta^{\prime}$ meson poles, with the vector bosons produced from $\pi^{0},\eta,\eta^{\prime}\rightarrow\gamma V,VV$.

\begin{figure}[t]
\centering      \includegraphics[width=15cm]{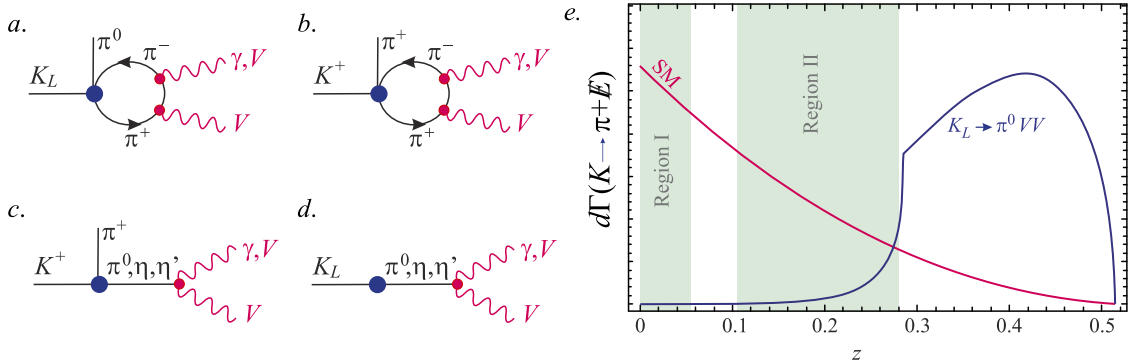}  \caption{$a.)$ In
chiral perturbation theory, the leading contribution to $K_{L}\rightarrow
\pi^{0}\gamma V,\pi^{0}VV$ arises from an $\mathcal{O}(p^{4})$ charged pion
loop. There is a similar loop contribution for the $K^{+}$ mode $(b.)$, but now
accompanied by a pole contribution $(c.)$ from the $\mathcal{O}(p^{4})$
odd-parity sector since $K^{+}$ is not a CP eigenstate. $d.)$ The
$K_{L}\rightarrow\gamma V,VV$ modes are purely induced by pole contributions
at leading $\mathcal{O}(p^{4})$ chiral order. $e.)$ The differential rate for
the pion loop-induced processes is strongly peaked above the $\pi\pi\pi$
threshold, hence falls out of the observation regions.}%
\label{FigDarkRad}%
\end{figure}

Let us now briefly discuss the simplest decay modes into dark vectors, and the corresponding bounds which could be set on $\varepsilon_{Q}$ and $\varepsilon_{\mathcal{B}}$. We start with the $K\rightarrow\pi V$ and $K\rightarrow\pi VV$ processes which would accompany $K\rightarrow\pi\nu\bar{\nu}$:

\begin{itemize}
\item $K\rightarrow\pi V$: For a massless vector, this process is forbidden (as is $K\rightarrow\pi\gamma$), so the dark vector boson has to be massive. At the same time, its mass should not be too large otherwise the pion momentum would fall out of the observation regions shown in Fig.~\ref{Kkin}. But even if it falls in the more favorable region II, $m_{\pi}<m_{V}<2m_{\pi}$, this process turns out to be very suppressed. It is directly related to $K\rightarrow\pi\gamma^{\ast}$, which does not arise at the leading chiral order~\cite{DEIP98}, and is insensitive to $\varepsilon_{\mathcal{B}}$. As a result, bounds on $\varepsilon_{Q}$ from $K^{+}\rightarrow\pi^{+}+\slashed E$ could at best be set at around $10^{-3}$. Note that $K_{L}\rightarrow\pi^{0}\gamma^{\ast}$ is CP-violating, hence not competitive compared to the charged mode (see Sec.~\ref{HadEff}).

\item $K_{L}\rightarrow\pi^{0}VV$: Being three-body, the vector mass should be small. When $m_{V}\ll m_{K}$, the rate is easily obtained from the corresponding photon mode as $\mathcal{B}(K_{L}\rightarrow\pi^{0}VV)\approx\varepsilon_{Q}^{4}\mathcal{B}(K_{L}\rightarrow\pi^{0}\gamma\gamma)$. But, with $\mathcal{B}(K_{L}\rightarrow\pi^{0}\gamma\gamma)^{\exp}=1.273(34)\times10^{-6}$~\cite{PDG}, a bound on $\mathcal{B}(K\rightarrow\pi VV)$ in the $10^{-12}$ range would at best translate as a bound on $\varepsilon_{Q}$ of the order of $5\%$. The situation is actually far worse than this, because this estimate disregards the important caution made in the previous section on the kinematics. Specifically, the $K_{L}\rightarrow\pi^{0}\gamma\gamma$ mode dominantly proceeds through a charged pion loop and its rate is significant only above the $\pi^{+}\pi^{-}$ threshold~\cite{ExpKpPpGG}. As shown in Fig.~\ref{FigDarkRad}, this is precisely the region excluded in the $K\rightarrow\pi+\slashed     E$ searches, because once $K_{L}\rightarrow\pi\pi\pi$ is open, one is swamped with backgrounds.

\item $K^{+}\rightarrow\pi^{+}VV$: This process is sensitive to both $\varepsilon_{Q}$ and $\varepsilon_{\mathcal{B}}$. The former arises from pion loops similar to those for $K_{L}\rightarrow\pi^{0}VV$. With $\mathcal{B}(K^{+}\rightarrow\pi^{+}\gamma\gamma)^{\exp}=1.10(32)\times10^{-6}$ and again a differential rate strongly peaked above the $\pi\pi\pi$ threshold, the bound on $\varepsilon_{Q}$ cannot compete with that from $K\rightarrow\pi V$. The pole contribution $K^{+}\rightarrow\pi^{+}(\pi^{0},\eta,\eta^{\prime}\rightarrow VV)$ offers in principle a window for $\varepsilon_{\mathcal{B}}$. But with $\mathcal{B}(K^{+}\rightarrow\pi^{+}\gamma\gamma)^{pole}$ presumably below about $10^{-7}$~\cite{GerardST05}, the sensitivity of $K^{+}\rightarrow\pi^{+}+\slashed      E$ on $\varepsilon_{\mathcal{B}}$ is not very good, and the other observables discussed below may offer better probes.
\end{itemize}

None of the $K\rightarrow\pi+\slashed  E$ modes appear very promising. Increasing the number of pions does not improve the sensitivity, since the branching ratios for the corresponding photonic modes are significantly smaller. Allowing for the presence of a photon, on the other hand, opens new interesting avenues to probe for the presence of a dark vector boson:

\begin{itemize}
\item $K\rightarrow\pi V\gamma$: At first sight, these modes may seem better candidates than $K\rightarrow\pi VV$ since $\varepsilon_{Q}$ or $\varepsilon_{\mathcal{B}}$ no longer appear to the fourth power. For example, with $\mathcal{B}(K_{L}\rightarrow\pi^{0}V\gamma)\approx\varepsilon_{Q}^{2}\mathcal{B}(K_{L}\rightarrow\pi^{0}\gamma\gamma)$, a bound in the $10^{-3}$ range on $\varepsilon_{Q}$ looks achievable. However, the differential rate is again similar to that of $K\rightarrow\pi\gamma\gamma$, see Fig.~\ref{FigDarkRad}. So, a bound in the $10^{-12}$ range for $K_{L}\rightarrow\pi\gamma+\slashed      E$ does not appear realistic, and no competitive bound on $\varepsilon_{Q}$ should be achievable. The situation is better for the pole contribution to $K^{+}\rightarrow\pi^{+}V\gamma$, sensitive to both $\varepsilon_{Q}$ and $\varepsilon_{\mathcal{B}}$, because it is not particularly suppressed at high pion momentum (low $z$ values). Using this mode, a bound at the percent level may be achievable.

\item $K_{L}\rightarrow\gamma V$: Finally, this is the best observable to look for dark vector bosons in the $K$ sector. First, it is induced by the anomaly, hence sensitive to both $\varepsilon_{Q}$ and $\varepsilon_{\mathcal{B}}$. Second, the probed vector mass range is the largest since there is no pion in the final state. Third, thanks to the presence of a photon, it is only quadratic in $\varepsilon_{Q}$ or $\varepsilon_{\mathcal{B}}$ instead of quartic. Fourth, the SM rate is not as precise as for $K\rightarrow\pi\nu\bar{\nu}$, but at $\mathcal{B}(K_L\rightarrow\gamma\nu\bar{\nu})\approx 3.4\times 10^{-13}$~\cite{KamenikS11}, it is sufficiently small not to obscure even a tiny NP contribution. Finally, with a bound at around $10^{-12}$, and with $\mathcal{B}(K_{L}\rightarrow\gamma\gamma)^{\exp}=5.47(4)\times10^{-4}$~\cite{PDG}, this mode could in principle probe $\varepsilon_{Q}$ and $\varepsilon_{\mathcal{B}}$ down to below $10^{-4}$, an order of magnitude better than any other observables in the few tens to few hundred MeV vector mass range.
\end{itemize}

All in all, the sensitivity of $K$ decays to a new flavor-blind vector boson suffers from the need to separately induce the flavor transition via the weak interaction. This is evident from the rather suppressed SM rates $\mathcal{B}(K_{L}\rightarrow\gamma\gamma)^{\exp}=5.47(4)\times10^{-4}$ and $\mathcal{B}(K\rightarrow\pi\gamma\gamma)^{\exp}$ around $10^{-6}$. Despite of this, the tremendous luminosity of the next generation of experiments, aiming for about a hundred of SM events for $K\rightarrow\pi\nu\bar{\nu}$, looks sufficient to competitively probe for dark vector bosons, especially if those couple predominantly to quarks. Dedicated strategies, in particular for the photon plus missing energy modes, should thus be designed and included in their experimental programs.

\section{Conclusion}

At this stage of the LHC program, the prospect for a new physics signal in the very rare $K$ decays may be dented, but remains well alive thanks to their intrinsic qualities. First, these decays are among the cleanest observables in the quark flavor sector. When combined with their terrible suppression in the SM, they thus offer uniquely sensitive probes. Second, the LHC capabilities are not ideal for all kinds of NP scenarios. For example, baryon number violating effects are often difficult to disentangle from the dominant QCD processes, while detecting a new very light but very weakly interacting particle is extremely challenging.

More generally, now that the simplest NP scenarios are in jeopardy, seemingly more exotic dynamics or particle contents may be our only alternative. Indeed, short of a profound reappraisal of our understanding of Nature, the hierarchy puzzle tells us that at least (some of) the lightest NP particles must be rather light, below or around the TeV, hence must have escaped detection at the LHC up to now. At the same time, such states may well induce deviations in flavor observables, since the dynamics of direct detection and of FCNC is not the same. So, and even though experimentalists should brace themselves for tiny deviations, rare $K$ decay have a clear role to play in the LHC era. This is perfectly illustrated by the Natural SUSY-like spectrum discussed in Sec.~\ref{NsusyMFV}. A relatively light stop can induce visible deviations in $K\rightarrow\pi\nu\bar{\nu}$, but it coexists with very heavy first and second generation squarks so as to avoid the direct detection bounds set at the LHC.

As a last message, let us stress once more an essential feature of kaon physics. Because its scale is deep in the non-perturbative QCD regime, predictions from first principle are not feasible. However, Nature has been kind enough to offer us the chiral symmetry. As any typical (spontaneously broken global) symmetry, it is usually powerless to predict a given observable, but instead permits to precisely relate observables among themselves. So, in contrast to other high-precision experiments concentrating on a single process like e.g. $\mu\rightarrow e\gamma$, EDM of some particle or nuclei, neutrinoless double beta transitions, or proton decay, a kaon physics experiment really benefits from an extensive experimental program. Of course, the very rare but very clean $K\rightarrow\pi\nu\bar{\nu}$ should have priority, but many other modes deserve special attention. Notably, as explained in Sec.~\ref{HadEff}, the semileptonic $K_{\ell 3}$ decays hold one of the keys to further improve the $K\rightarrow\pi\nu\bar{\nu}$ predictions, while radiative decays are crucial to control various sources of theoretical uncertainties for the FCNC processes involving charged leptons like $K_L\rightarrow\mu^+\mu^-$, $K_L\rightarrow\pi^0e^+e^-$, and $K_L\rightarrow\pi^0\mu^+\mu^-$. By the way, those should certainly not be disregarded. Compared to $K\rightarrow\pi\nu\bar{\nu}$, they have complementary sensitivities to NP. In addition, as mentioned in Sec.~\ref{portals}, a dedicated physics program for radiative decays offer the possibility to competitively probe for new exotic and very light forms of matter or interactions.

In conclusion, kaon physics retains its peculiar place in the LHC era. After more than half-a-century of good service, and along the way, the game-changing discoveries of indirect and direct CP violation, it may still rise to preeminence thanks to the exceptional theoretical cleanness of the rare $K$ decays and the incredible experimental luminosity of the planned experiments.

\subsection*{Acknowledgements}

Many thanks to the organizers of FPCP 2014 for their kind invitation.


\begin{thebibliography}{99}                                                                                                

\bibitem {BuchallaBL96}G.~Buchalla, A.~J.~Buras and M.~E.~Lautenbacher, Rev. Mod. Phys. \textbf{68} (1996) 1125.

\bibitem {BurasZp}For a recent explicit realization of this kind of scenarios,
see e.g. A.~J.~Buras, D.~Buttazzo, J.~Girrbach-Noe and R.~Knegjens, arXiv:1408.0728 [hep-ph].

\bibitem {MFV}G.~D'Ambrosio, G.~F.~Giudice, G.~Isidori and A.~Strumia, Nucl.\ Phys.\ B \textbf{645} (2002) 155;
See also e.g. L.~J.~Hall and L.~Randall, Phys.\ Rev.\ Lett.\ \textbf{65} (1990) 2939;
A.~Ali and D.~London, Eur.\ Phys.\ J.\ C \textbf{9} (1999) 687;
A.~J.~Buras, P.~Gambino, M.~Gorbahn, S.~Jager and L.~Silvestrini, Phys.\ Lett.\ B \textbf{500} (2001) 161;
C.~Smith, Acta Phys. Polon. Supp. \textbf{3} (2010) 53.

\bibitem {BurasG14}For a recent review, see e.g. A.~J.~Buras and J.~Girrbach, Rept.\ Prog.\ Phys.\ \textbf{77} (2014) 086201.

\bibitem {InamiLim}T.~Inami and C.~S.~Lim, Prog.\ Theor.\ Phys.\ \textbf{65} (1981) 297 [Erratum-ibid.\ \textbf{65} (1981) 1772].

\bibitem {GIM}S.~L.~Glashow, J.~Iliopoulos and L.~Maiani, Phys.\ Rev.\ D \textbf{2} (1970) 1285.

\bibitem {kpnnLD}D.~Rein and L.~M.~Sehgal, Phys.\ Rev.\ D \textbf{39} (1989) 3325;
J.~S.~Hagelin and L.~S.~Littenberg, Prog.\ Part.\ Nucl.\ Phys.\ \textbf{23} (1989) 1;
M.~Lu and M.~B.~Wise, Phys.\ Lett.\ B \textbf{324} (1994) 461;
G.~Buchalla and G.~Isidori, Phys.\ Lett.\ B \textbf{440} (1998) 170.

\bibitem {IsidoriMS05}G. Isidori, F. Mescia, C. Smith, Nucl. Phys. B \textbf{718} (2005) 319.

\bibitem {kpnnSD}A.~J.~Buras, M.~Gorbahn, U.~Haisch and U.~Nierste, Phys. Rev. Lett. \textbf{95} (2005) 261805;
JHEP \textbf{0611} (2006) 002;
J.~Brod and M.~Gorbahn, Phys.\ Rev.\ D \textbf{78} (2008) 034006.

\bibitem {BGS11}J.~Brod, M.~Gorbahn and E.~Stamou, Phys.\ Rev.\ D \textbf{83} (2011) 034030.

\bibitem {BB96}G. Buchalla, A.J. Buras, Phys.\ Rev.\ D \textbf{54} (1996) 6782.

\bibitem {GrossmanN97}Y.~Grossman and Y.~Nir, Phys.\ Lett.\ B \textbf{398} (1997) 163.

\bibitem {DEIP98}G.~D'Ambrosio, G.~Ecker, G.~Isidori and J.~Portoles, JHEP \textbf{9808} (1998) 004.

\bibitem {BDI03}G.~Buchalla, G.~D'Ambrosio and G.~Isidori, Nucl. Phys. B \textbf{672} (2003) 387.

\bibitem {ISU04}G.~Isidori, C.~Smith and R.~Unterdorfer, Eur. Phys. J. C \textbf{36} (2004) 57.

\bibitem {MesciaS07}F.~Mescia and C.~Smith, Phys.\ Rev.\ D \textbf{76} (2007) 034017;
J.~Bijnens, K.~Ghorbani, arXiv:0711.0148 [hep-ph].

\bibitem {BGB14}A.~J.~Buras, J.~M.~G\'{e}rard and W.~A.~Bardeen, Eur.\ Phys.\ J.\ C \textbf{74} (2014) 2871.

\bibitem {FGD04}C.~Bruno, J.~Prades and Z.\ Phys.\ C\textbf{ 57} (1993) 585;
S. Friot \textit{et al.}, Phys. Lett. B \textbf{595} (2004) 301.

\bibitem {NA48as}J.~R.~Batley \textit{et al.} [NA48/1 Collaboration], Phys.\ Lett.\ B \textbf{576} (2003) 43;
Phys.\ Lett.\ B \textbf{599} (2004) 197.

\bibitem {KLpgg}A.~Lai \textit{et al.} [NA48 Collaboration], Phys.\ Lett.\ B \textbf{536} (2002) 229;
E.~Abouzaid \textit{et al.} [KTeV Collaboration], Phys.\ Rev.\ D \textbf{77} (2008) 112004.

\bibitem {MertensS11}P.~Mertens and C.~Smith, JHEP \textbf{1108} (2011) 069.

\bibitem {KPpnunu}S.~S.~Adler \textit{et al.} [E787 Collaboration], Phys. Rev. Lett. \textbf{88} (2002) 041803;
A.~V.~Artamonov \textit{et al.} [E949 Collaboration], Phys. Rev. Lett. \textbf{101} (2008) 191802.

\bibitem {K0pnunu}J.~K.~Ahn \textit{et al.} [E391a Collaboration], Phys. Rev. D \textbf{81} (2010) 072004.

\bibitem {KTeVelec}A.~Alavi-Harati \textit{et al.} [KTeV Collaboration], Phys. Rev. Lett. \textbf{93} (2004) 021805.

\bibitem {KTeVmuon}A.~Alavi-Harati \textit{et al.} [KTEV Collaboration], Phys.\ Rev.\ Lett.\ \textbf{84} (2000) 5279.

\bibitem {MesciaST}F. Mescia, C. Smith, S. Trine, JHEP \textbf{0608} (2006) 088.

\bibitem {IsidoriU04}G.~Isidori, R.~Unterdorfer, JHEP \textbf{0401} (2004) 009.

\bibitem {DAmbrosioP97}G.~D'Ambrosio and J.~Portoles, Nucl.\ Phys.\ B \textbf{492} (1997) 417.

\bibitem {DAmbrosioIP98}G.~D'Ambrosio, G.~Isidori and J.~Portoles, Phys.\ Lett.\ B \textbf{423} (1998) 385.

\bibitem {GerardST05}J.-M.~G\'{e}rard, C.~Smith, S.~Trine, Nucl. Phys. \textbf{B730} (2005) 1.

\bibitem {CappielloCDG12}L.~Cappiello, O.~Cata, G.~D'Ambrosio and D.~N.~Gao, Eur.\ Phys.\ J.\ C \textbf{72} (2012) 1872 [Erratum-ibid.\ C \textbf{72}
(2012) 2208].

\bibitem {NSUSY}A.~G.~Cohen, D.~B.~Kaplan and A.~E.~Nelson, Phys.\ Lett.\ B \textbf{388} (1996) 588;
R.~Kitano and Y.~Nomura, Phys.\ Rev.\ D \textbf{73} (2006) 095004;
R.~Barbieri and D.~Pappadopulo, JHEP \textbf{0910} (2009) 061;
M.~Papucci, J.~T.~Ruderman and A.~Weiler, JHEP \textbf{1209} (2012) 035.

\bibitem {Chargino}Y. Nir and M.P. Worah, Phys.\ Lett.\ B \textbf{423} (1998) 319;
A.J. Buras, A. Romanino, L. Silvestrini, Nucl.\ Phys.\ B \textbf{520} (1998) 3;
A. J. Buras, G. Colangelo, G. Isidori, A. Romanino and L. Silvestrini, Nucl. Phys. B \textbf{566} (2000) 3.

\bibitem {ColangeloI}G. Colangelo and G. Isidori, JHEP \textbf{9809} (1998) 009.

\bibitem {IMPST06}G.~Isidori, F.~Mescia, P.~Paradisi, C.~Smith and S.~Trine, JHEP \textbf{0608} (2006) 064.

\bibitem {MFVapp}A. J. Buras, P. Gambino, M. Gorbahn, S. Jager, L. Silvestrini, Nucl. Phys. B \textbf{592} (2001) 55;
C.~Bobeth \emph{et al.}, hep-ph/0505110.

\bibitem {NSUSYMFV}F.~Br\"{u}mmer, S.~Kraml, S.~Kulkarni and C.~Smith, arXiv:1402.4024 [hep-ph].

\bibitem {MFVNiko}E.~Nikolidakis, Ph.D. thesis, University of Bern (2008).

\bibitem {Georgi}R.~S.~Chivukula and H.~Georgi, Phys.\ Lett.\ B \textbf{188} (1987) 99.

\bibitem {MFVMercolli}L.~Mercolli and C.~Smith, Nucl.\ Phys.\ B \textbf{817} (2009) 1.

\bibitem {MFVrunning}P.~Paradisi, M.~Ratz, R.~Schieren and C.~Simonetto, Phys.\ Lett.\ B \textbf{668} (2008) 202;
G.~Colangelo, E.~Nikolidakis and C.~Smith, Eur. Phys. J. C \textbf{59} (2009) 75.

\bibitem {MFVRPV}E.~Nikolidakis and C.~Smith, Phys.\ Rev.\ D \textbf{77} (2008) 015021;
C.~Smith, Talk given at ICHEP08, Philadelphia, PA, 30 Jul - 5 Aug 2008, arXiv:0809.3152 [hep-ph].

\bibitem {Barbier04}For a review of R-parity violation, see e.g. R.~Barbier \textit{et al.}, Phys.\ Rept.\ \textbf{420} (2005) 1.

\bibitem {MFVRPVCGH}C.~Csaki, Y.~Grossman and B.~Heidenreich, Phys.\ Rev.\ D \textbf{85} (2012) 095009.

\bibitem {MFV4G}C.~Smith, Phys.\ Rev.\ D \textbf{85} (2012) 036005.

\bibitem {tHooft76}G.~'t Hooft, Phys.\ Rev.\ Lett.\ \textbf{37} (1976) 8;
Phys.\ Rev. D\ \textbf{14} (1976) 3432 [Erratum-ibid. D \textbf{18} (1978) 2199].

\bibitem {MFVRPVrun}J.~Bernon and C.~Smith, JHEP \textbf{1407} (2014) 038.

\bibitem {MFVLHC1}G.~Durieux, J.~M.~G\'{e}rard, F.~Maltoni and C.~Smith, Phys.\ Lett.\ B \textbf{721} (2013) 82.

\bibitem {MFVLHC2}G.~Durieux and C.~Smith, JHEP \textbf{1310} (2013) 068.

\bibitem {BuchmullerW86}W.~Buchmuller and D.~Wyler, Nucl.\ Phys.\ B \textbf{268} (1986) 621.

\bibitem {Weinberg}S.~Weinberg, Phys.\ Rev.\ Lett.\ \textbf{43} (1979) 1566.

\bibitem {JaeckelR10}J.~Jaeckel and A.~Ringwald,
Ann.\ Rev.\ Nucl.\ Part.\ Sci.\ \textbf{60} (2010) 405.

\bibitem {WeinbergA}S.~Weinberg, Phys.\ Rev.\ Lett.\  {\bf 40} (1978) 223.

\bibitem {KamenikS11}J.~F.~Kamenik and C.~Smith, JHEP \textbf{1203} (2012) 090.

\bibitem {Neutralinos}H.~K.~Dreiner, S.~Grab, D.~Koschade, M.~Kramer, B.~O'Leary and U.~Langenfeld, Phys.\ Rev.\ D \textbf{80} (2009) 035018.

\bibitem {Fayet89}P.~Fayet, Phys.\ Lett. B\ \textbf{227}(1989) 127.

\bibitem {Essig}R.~Essig, P.~Schuster, N.~Toro and B.~Wojtsekhowski, JHEP \textbf{1102} (2011) 009;
R.~Essig, R.~Harnik, J.~Kaplan and N.~Toro, Phys.\ Rev.\ D \textbf{82} (2010) 113008.

\bibitem {Williams}M.~Williams, C.~P.~Burgess, A.~Maharana and F.~Quevedo, JHEP \textbf{1108} (2011) 106.

\bibitem {Heeck}J.~Heeck, arXiv:1408.6845 [hep-ph].

\bibitem {PDG}K. Nakamura \textit{et al.} [Particle Data Group], J. Phys. G \textbf{37} (2010) 075021.

\bibitem {ExpKpPpGG}C.~Lazzeroni \textit{et al.} [NA62 Collaboration], Phys.\ Lett.\ B \textbf{732} (2014) 65.

\end{thebibliography}
\end{document}